\documentclass[reprint,
superscriptaddress,
 amsmath,amssymb,
 aps,
pra,
]{revtex4-2}

\usepackage{graphicx}% Include figure files
\usepackage{dcolumn}% Align table columns on decimal point
\usepackage{xcolor}
\usepackage{bm}% bold math
\usepackage{physics}
\usepackage{hyperref}
\usepackage{algorithm}
\usepackage{algpseudocode}
\hypersetup{
	colorlinks = true,
	urlcolor   = blue,
	linkcolor  = blue,
	citecolor  = red
}
\begin{document}

\title{
%Genetic algorithm for 
Identifying network topologies %of  graphs 
via quantum walk distributions }
% Force line breaks with \\

\author{Claudia Benedetti}
\affiliation{Dipartimento di Fisica ``Aldo Pontremoli'', Universit\`a degli Studi di Milano, 20133, Milan, Italy}

\author{Ilaria Gianani}
\affiliation{Dipartimento di Scienze, Universit\`{a} degli Studi Roma Tre, Via della Vasca Navale 84, 00146, Rome, Italy}
 \email{ilaria.gianani@uniroma3.it}

\begin{abstract}
Control and characterization of networks is a paramount step for the development of many quantum technologies. Even for moderate-sized networks, this amounts to explore an extremely vast parameters space in search for the couplings defining the network topology. Here we explore the use of a genetic algorithm to retrieve the topology of a network from the measured probability distribution obtained from the evolution of a continuous-time quantum walk on the network. Our result shows that the algorithm is capable of efficiently retrieving the required information even in the presence of noise. 
\end{abstract}

\maketitle
%\begin{itemize}
%    \item CN in QT
%    \item Simulating large networks
%    \item Simulation paradigm 
%    \item Problem: assess topology from QW distribution  (stress importance of accessing info through easily measured data) 
 %   \item Here we solve it using a genetic algorithm 
%\end{itemize}

Networks are a fundamental model to understand the underlying properties of complex systems. They are invaluable tools to describe phenomena happening at different scales ranging 
from social interactions \cite{wasserman94,Onnela07}, to  biological processes \cite{jeong00,pastor01, maslov02,silva05,plenio08}, from the configurations of molecules \cite{Winterbach13,dekeer21}, to the structure of internet \cite{faloutsos99, Caldarelli_2000,Pastor04, He2009} and physical systems alike \cite{zoller97,kuzmich05,mulken16,krutitsky16,nokkala16}. 
In the context of quantum technologies, networks constitute the prime structure of communication and computation protocols \cite{deutch89,Christandl05,bose07,politi08,aspuru12}. 
Understanding how quantum information can be reliably transmitted between distant nodes of a network, or routed among different computational units, is a key step and requires a full characterization of the network's structure.  
While a direct control may not be attainable with the required accuracy and precision, a straightforward strategy to provide such characterization is that of probing the network with a walker that gathers information on its topology by undergoing an evolution which depends on the network's structure. 
This is the case of continuous-time quantum walks (CTQWs) \cite{fahri98,childs_cleve_03,kendon2006quantum,kay10,mulken11,benedetti19,Chakraborty20,gualtieri20,kadian21,Bressanini22}, which thus emerge as a natural paradigm for tackling this task.

Two different scenarios may present: the topology of the network may be known, but an accurate estimation of the coupling strengths between each node may be required. This is tantamount to estimating multiple parameters, and can be address in quantum metrological terms \cite{gianani22, tama16, Seveso_2019,annabestani22}. 
It might otherwise be the case that the topology of the network is not known in advance. 
Whether one is interested in characterizing a physical network or a simulated one, this will be relying on an experimental platform controlled with set of experimental parameters $\mathbf{\Lambda^{exp}}$. 
These need to be mapped to the associated set of parameters describing  the CTQW happening on the network, $\mathbf{\Lambda^{QW}}$, i.e. the Hamiltonian parameters of the quantum walk which, assuming all coupling strengths are fixed to unity and on-site energies to zero, coincide with the adjacency matrix identifying the topology of the network. 
In order to asses the evolution of the probe, one has to address an observable, such as the spatial probability distribution on the network, which will strongly depend on the network's topology. 
However, since an analytical description of this distribution for CTQWs is often unattainable, and furthermore the relation between the QW Hamiltonian and its probability distribution is highly non-linear, performing a direct inversion can be involved. 
At the same time, the parameter space in this instance becomes exceedingly large for this to be treated as an estimation problem. 
An alternative solution is to cast the 
issue in terms of a search problem. Having access solely to the initial state of the probe and to the measured experimental distributions at fixed times, the task becomes that of finding an adjacency matrix that matches the evolution. 
Here we tackle this matter by using a genetic algorithm. We use the algorithm to successfully retrieve different topologies in the ideal case as well as when the measured probabilities are affected by noise. 

We consider a CTQW with zero on-site energies, defined by the couplings $\mathbf{\Lambda^{QW}}=\{J_{xy}\}$ between two nodes of the network $x$ and $y$, such that its Hamiltoian is: 
\begin{equation}
H(\mathbf{\Lambda^{QW}})=\sum_{xy}J_{xy}\ketbra{x}{y}.
\label{hami}
\end{equation}
We assume that the couplings $J_{xy}$ can take only two values: $J_{xy}=0$ if the link between two nodes is off, or $J_{xy}=1$ if the link is on, so that each edge is bound to have the same strength. The Hamiltonian thus coincides with the adjacency matrix of the network, hence, determining its parameters amounts to determining the network's topology. The evolution of a walker in the initial state $\ket{\psi_0}$ is described by the unitary  operator $e^{-i H t}$. The probability of occupying a site $x$ at a time $t$ is then: 
 \begin{align}
p_x(t,\mathbf{\Lambda^{QW}})=|\bra{x}e^{-i H(\mathbf{\Lambda^{QW}}) t}\ket{\psi_0}|^2.
\label{probx}
 \end{align}
Given an {undirected graph} of n sites, our objective is that of retrieving the couplings $\mathbf{\Lambda^{QW}}=\{J_{12},\dots,J_{(\text{n}-1)\text{n}}\}$, i.e. a binary string of length $n_c=$n$($n$-1)/2$, having access only to the initial state of the network and to the probabilities $p_x(t_k,\mathbf{\Lambda^{QW}})$ measured at times $t_k$. 
 \begin{figure*}
    \includegraphics[width=1\textwidth]{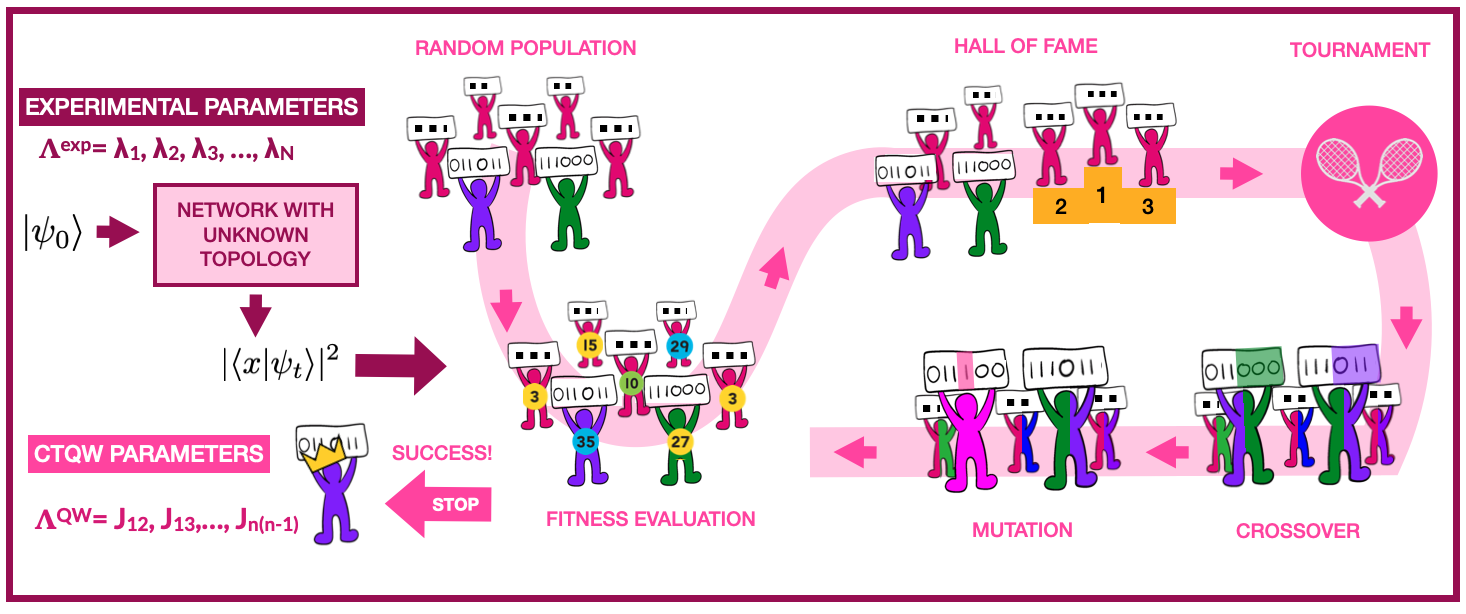}
    \caption{{\bf Conceptual scheme.}  Given an initial probe state $\vert \psi_0\rangle$ and a network with unknown topology controlled by a set of experimental parameters, we aim at retrieving the topology of the network measuring the probability distributions of the the probe evolved with a CTQW. This is achievede through a genetic algorithm in which the probability distributions are employed to evaluate the fitness score, as described in the main text. }
    \label{fig:scheme}
\end{figure*}  
We tackle this challenge by means of a genetic algorithm (GA). GAs are versatile iterative search algorithms inspired by natural selection and have been extensively employed for quantum tasks \cite{Rambhatla20,Knott2016,Nichols2019,Lukac2003}. 
They rely on the evolution of a population of individuals, each defined by a chromosome string and a fitness score, which breed new individuals replacing the previous population at each iteration. By promoting the reproduction of the fittest individuals while introducing various mechanisms to ensure enough genetic variability, GAs allow to efficiently retrieve the optimal solution\cite{mitchell98,katoch21}. 

We encode the chromosomes as binary strings $\Lambda_i$ of length $n_c$, so that each gene constituting the chromosome is a coupling $J_{xy}$. 
The fitness of each individual is evaluated as follows: 
$\Lambda_i$ is used to evolve the initial state of the probe up to selected times $t_k$ obtaining the probability distributions $p_x(t_k, \Lambda_i)$. For practical purposes, we concatenate the  probabilities  at different times in a single array that we call $\pi_x(\{t_k\},\Lambda_i)$.
Using multiple times allows to remove eventual ambiguities and to mitigate the effects of local minima, thus improving the performance of the algorithm. 
We  then check the distance between these probabilities and the  measured ones $\pi_x(\{t_k\}, \mathbf{\Lambda^{QW}})$ e.g. by using the Kullback-Leibler divergence. When the distance is null, $\Lambda_i=\mathbf{\Lambda^{QW}}$. The value of the distance will be the fitness score of each individual. 
Thus, in our case, the more fit an individual, the smaller its fitness score. The correct couplings will be those having a fitness score equal to 0.

The algorithm scheme is shown in  Fig. \ref{fig:scheme} and operates as follows: 
An initial random population of size $n_p$ is generated, and its fitness is evaluated as described above. An {\it elitist} function selects a small percentage $p_e$ of individuals with the best fitness scores to constitute the {\it hall of fame},  which will be cloned in the next generation. The whole population is then entered in a tournament where $k$ individuals at the time compete to be selected for breeding the next generation. This is achieved through a crossover strategy in which the chromosomes of the selected parents are mixed with a probability $p_c$. The size of the population is kept constant through each generation, so that each selected pair of individuals will produce two children. In order to ensure genetic diversity, with a small probability $p_m$, children can undergo mutations, consisting in bit flips. The new born children together with the cloned hall of fame individuals become the next generation, and the algorithm continues iteratively, stopping either when the optimal individual is found (i.e. fitness score equals to zero), or when a maximum number of generations $n_g$ is reached. A full depiction of the algorithm and of the implementation of the genetic operations are reported in the Supplementary Information. 

 \begin{figure*}[t!]
    \includegraphics[width=1\textwidth]{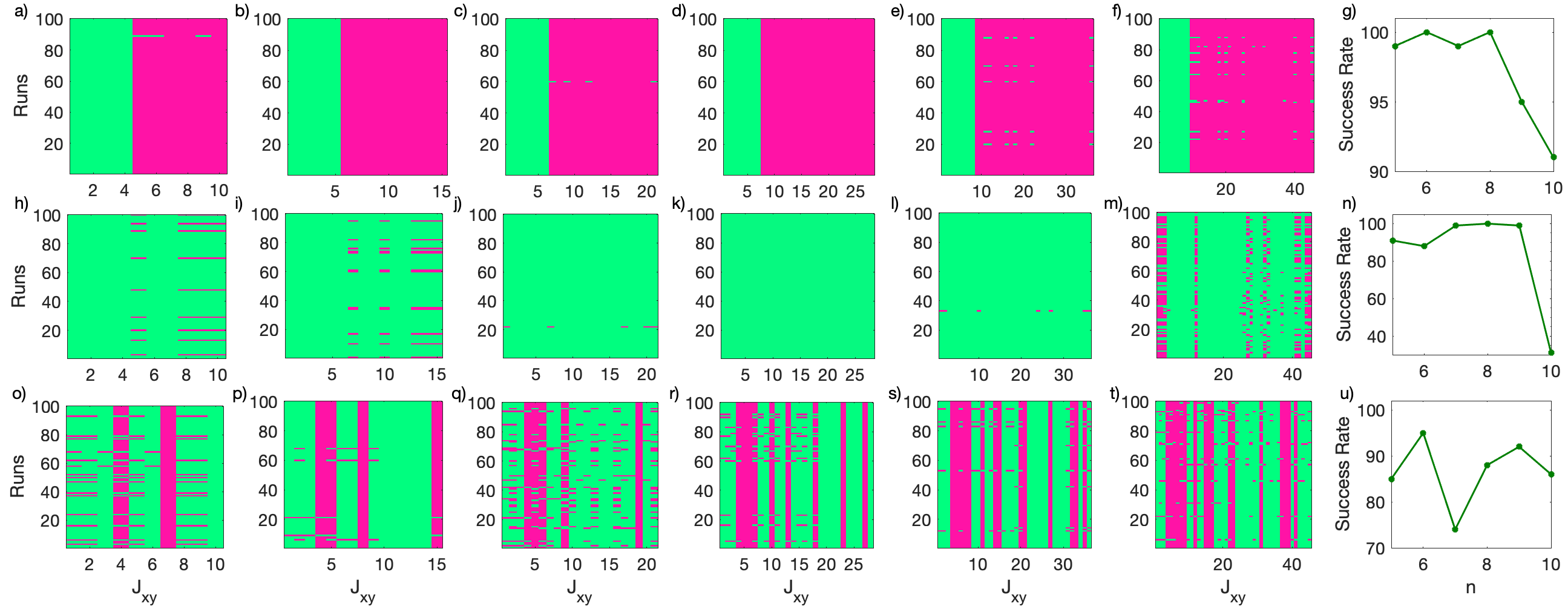}
    \caption{{\bf Results without noise} Retrieved couplings for N runs of the algorithm for star (a-f), complete (h-m), and random (o-t) networks for varying network sizes: (a,h,o) n=5, (b,i,p) n=6, (c,j,q) n=7, (d,k,r) n=8, (e,l,s) n=9 (f,m,t) n=10. (g,n,u) success rates as a function of network size. Green indicates a coupling equal to 1, fuchsia a coupling equal to 0.}
    \label{fig:couplings}
\end{figure*}  

The initial state of the probe as well as the evolution times at which the probabilities are measured play a fundamental role towards the success of the algorithm: for instance, choosing a localized state may result in discarding part of the network, if composed by two or more disjoints subnetworks; evolving the state for too short a time in a large network, may preclude the state to reach the whole network. 
While we do not perform a full optimization of the initial state, we  choose one that allows to explore a large variety of different topologies and network sizes. Also all the hyperparameters defining the algorithm (population size $n_p$, elitist population $p_e$, individuals involved in each tournament $k$, crossover probability $p_c$, mutation probability $p_m$, max number of generations $n_g$) can be optimized in accordance with the task at hand and specifically with the network size. 
In our analysis we vary the network size to explore how the algorithm scales with an increasing number of couplings, but for the sake of simplicity we have chosen to keep all hyperparameters fixed aside from the population size $n_p$. 
Our results hence are but a lower bound to the achievable performance attainable by fine-tuning for a fixed network size. 

Here we report the results obtained with a star graph, a complete graph  and a graph with an arbitrary topology. 
This last network is a simplified version  of the graph in Ref \cite{knuth} describing the relations between the characters in the novel Les Misérables \cite{hugo}. 
Results for additional topologies (line and circle) and further details on the generation of the Les Misérables graph  can be found in the Supplementary Information.  
\begin{figure}[t!]    \includegraphics[width=1\columnwidth]{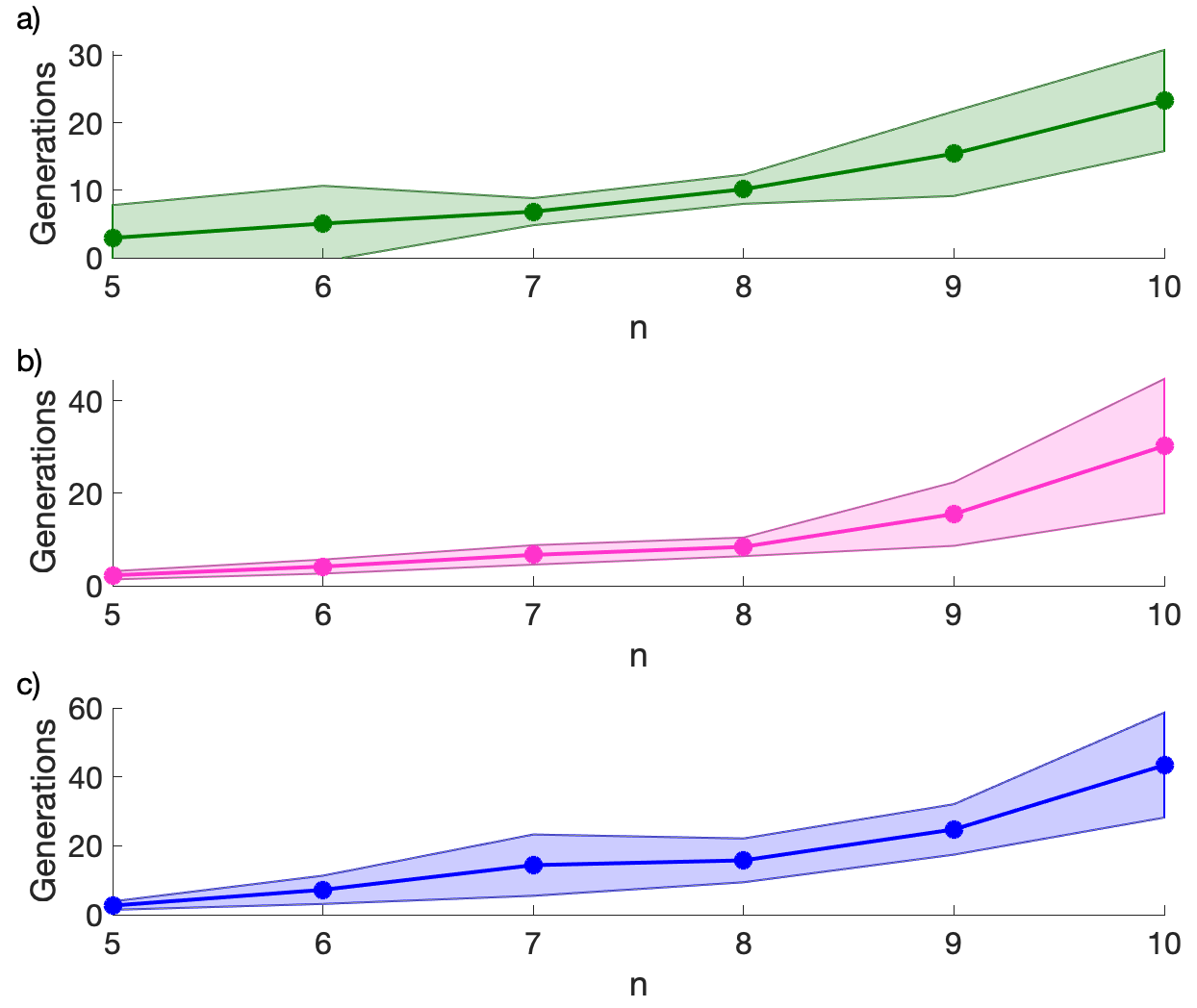}
\caption{{\bf Algorithm convergence.} Average numbers of generations required for convergence over N runs of the algorithm for a star (a), complete (b), and random (c) network as described in the main text. The shaded region is the standard deviation error over the N runs.}
\label{fig:gen}
\end{figure}  
In order to test the algorithm, we inspect networks with nodes from n=5 to n=10, thus we search for binary strings with length $n_c=10$ to $n_c=45$. We measure the probability distributions at two distinct times, $t_1=0.5$ and $t_2=0.6$. 
As mentioned above, all hyperparameters are kept fixed (see Supplementary Information), aside from the population size $n_p$ which we scale as $n_p=2\cdot n_c^2$. This ensures a trade-off between computation time and performance, and allows us to provide a controlled comparison for the performance at different sizes. We fix the maximum number of generations to $n_g=100$, and, for each configuration, we run the algorithm N=100 times. 

We first consider the ideal case in which the probability distributions are noiseless. The results are reported in Fig \ref{fig:couplings}, which shows the couplings values (green = 1, fuchsia=0) obtained for each run of N for the star  (a-f), complete  (h-m) and Les Misérables graph (o-t), as well as the success rate in each instance (g, n, u). 
Fig. \ref{fig:couplings} highlights how most of the times when the algorithm fails it returns the same (wrong) couplings. 
This effect is due to the algorithm getting stuck in the same local minima because for the chosen evolution times  there are multiple configurations leading to probabilities which are very close to the true one. 
The most affected network is the complete, whose success rate, for n=10, drops to $31\%$. However, it is sufficient to run the algorithm including also a third probability measured at time $t_3=1$, and a success rate of $73\%$ is recovered (see the Supplementary Information). 
In Fig. \ref{fig:gen} we report the number of generations needed for convergence as a function of n, which predictably increases with the number of network sites, as does the search space. 
Our results show a remarkable efficiency of the search algorithm employed: in fact, the number of possible combinations $\Lambda_i$ scales with $2^{n_c}$, while we are inspecting, at most, $2\cdot n_c^2\cdot n_g$ combinations, assuming the worst case scenario in which we run the algorithm for the maximum number of generations and completely replace the population each time. For n=10, when $n_c=45$ the combinations hence amount to $\sim 3.5\cdot 10^{13}$, and we are exploring less than $\sim 4\cdot 10^5$ configurations. 
\begin{figure}[t!]
\includegraphics[width=1\columnwidth]{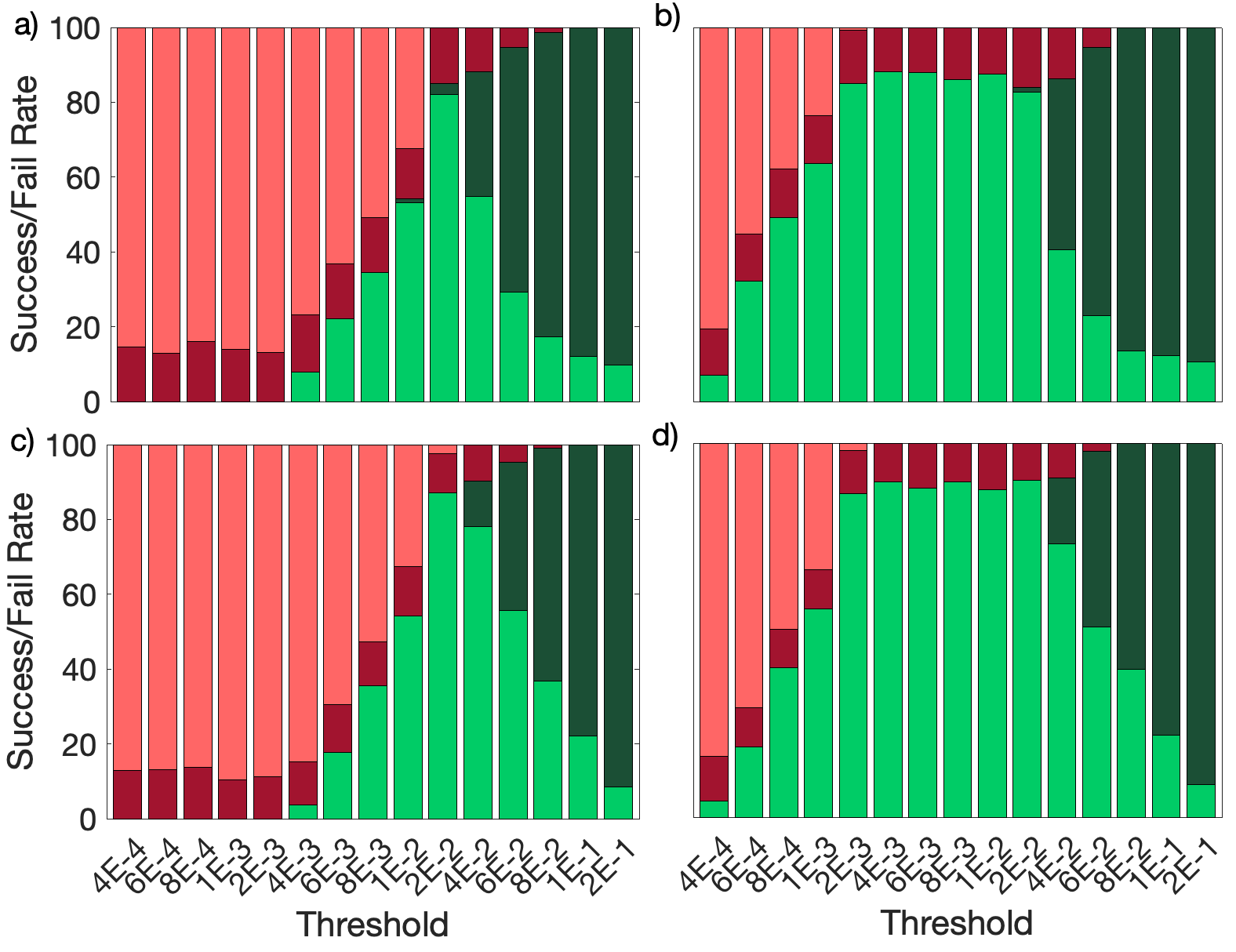}
\caption{{\bf Results with noise.} Success/fail rates over the MC runs for the algorithm performed on a star network (a-b) and a complete network (c-d) for $N_r=500$ resources (a-c) and $N_r=5000$ resources (b-d). Light green: true positives;  Dark green: false positives; Light red: false negatives; Dark red: true negatives.}
\label{fig:noise}
\end{figure}  

In a real-life scenario, the probabilities $p(t_k,\mathbf{\Lambda^{QW}})$ used to evaluate the fitness score would be affected by noise. This needs to be accounted for when evaluating the distance by setting a threshold value T below which  two probabilities are considered equal. The algorithm thus needs to be modified to halt whether the distance between the measured and evaluated probabilities is smaller than  T , which counts as a success, or when it reaches the maximum number of generations, in which case the algorithm has failed. Depending on the value of T, there can be four outcomes:
1) {\it True negative}: The algorithm fails to reach T and the couplings are not found. 
2) {\it False negative}: The algorithm fails to reach T, but the exact string of couplings has been found. This happens if T is set too low compared to the noise affecting the probabilities. 
3) {\it True positive}: the alogirhtm successfully finds a fitness below T, and that corresponds to the exact couplings.
4) {\it False positive}: the algorithm successfully finds a fitness below T, but the couplings are not correct. This happens when the threshold is set too high compared to the noise, and hence the algorithm stops before it can converge. 
In order to test this behaviour, we simulate the measured probabilities for a network with n=5 for a star topology and a fully connected topology, using the same hyperparameters as before aside from the max number of generations which we set to $n_g=5$.
We know from the ideal case (Fig. \ref{fig:gen}) that for these topologies the algorithm converges in more than 5 generations, so we do expect to have some true negative outcomes.
We simulate the probability measurements with a total of $N_r$ resources  ranging from $N_r=500$ to $N_r=5000$, and through a Montecarlo (MC) routine we add Possionian noise to the simulated probabilities. 
For each MC run, we average the successes/fails over N=10 runs of the algorithm. We record the results for threshold values ranging from T$=4\cdot10^{-4}$ to T$=0.2$. 
In Fig. \ref{fig:noise} we report the results of the success/fail rates over 100 MC runs for the star network (a-b) and complete network (c-d) with $N_r=500$ (a-c) and $N_r=5000$ (b-d) as a function of the threshold value (for other $N_r$ see Supplementary Information). 
As expected, we can observe the four behaviours described early: when T is too low, the outcomes are dominated by false negatives (light red), with a small percentage of true negatives, due to the fact that the algorithm would take more than 5 generations to converge. As the threshold increases so do the number of true positives, while the true negatives remain constant. 
For larger thresholds both the true positive and true negative drop. 
The algorithm always satisfies the threshold condition before it can converge to the actual solution. 

 In conclusion, we have employed a genetic algorithm to retrieve the topology of a network having access solely to the initial state of a probe undergoing a CTWQ and to the measured probability distributions at given times. 
 We have explored the performance of the algorithm for different network sizes and topologies, as well as when the measured probabilities are affected by Poissonian noise. The algorithm maintains high performance levels for all the configurations explored, which could be further optimized by fine-tuning the hyperparameters for a specific network size. The genetic algorithm is particularly suited to address large parameter spaces, however increasing the network size by order of magnitudes or remove the constraint on the coupling strength would make it challenging in terms of computational times and resources. In order to achieve such scalability, a perspective is that of extending these results to incorporated machine learning techniques. 
 By relying solely on measured probabilities, our technique provides  a simple but yet effective strategy for the routine characterization of networks, and as such constitutes an enabling step towards most developing quantum technologies based on complex networks \cite{briegel98,kimble08,cirac10,benedetti21,coutinho22,candeloro22}, as well as a tool for exploring new involved simulation regimes which have non-trivial mapping between the experimental control and the CTQW parameters \cite{karski09,imany20,kaufman22}.

\bibliography{main.bib}% Produces the bibliography via BibTeX.

\section{Appendix}

\subsection{genetic algorithm}

The algorithm begins with an initial population initialized by generating $n_p$ random binary arrays $\Lambda_i$ of size $n_c$, containing the couplings $J_{xy}$, which in this representation correspond to the genes of each individual. 
These $n_p$ chromosomes correspond to the zeroth generation. 
While the number of generations is lower than $n_g$, we proceed as follows: 
We evaluate the score $S_i$ of each string $\Lambda_i$ using the Fitness function described in details in the next section. 
The best fitness score, corresponding to the lowest value, and the relative couplings are stored. 
If the score is equal to zero, the optimal solution has been found, the algorithm stops and returns it.
If the condition is not met, the algorithm continues by selecting the fittest $p_e\cdot n_p$ chromosomes $\Lambda_i$ and places them in a hall of fame, to be cloned in the following generation.
Since the population size has to stay constant,  we  need to create the remaining $n_p(1-p_e)$ individuals which will populate the next generation together with those cloned from the hall of fame. 
In order to do so we  select the best parents from the whole population (including the hall of fame). This is achieved with the Tournament selection function, which randomly selects $k$ individuals at a time and returns the best among them (lowest fitness score). 
The random selection of the $k$ competitors  ensures that the chosen individuals are not necessarily the best in the population. In this way, genetic diversity is ensured to mitigate the presence of local minima. Once the parents are selected, they are mixed through the Crossover function, which returns two children which, with probability $p_c$, are composed by a mixture of the parents chromosomes.
To further ensure genetic diversity, the genes of the children can undergo mutations with mutation probability $p_m$. When a mutation happens, the gene is flipped. 
The generated children together with the hall of fame constitute the new generation. The algorithm repeats until either a chromosome with fitness score equal to zero is found, or the maximum number of generations is reached. The pseudocode of the algorithm reported in Algorithm \ref{genalg}.

\begin{algorithm}[H]
\caption{Genetic Algorithm}\label{genalg}
\begin{algorithmic}[1]
\State $gen\leftarrow 0$
\State    Randomly generate $n_p$ binary arrays $\{\Lambda_i\}$
\State $P_{gen}\,\leftarrow\{\Lambda_i\}$
\Comment{Initialize population}
\While {$gen<n_g$}
\For {$i=0 \rightarrow  n_p-1$}
\State $S_i$= \Call{Fitness}{$\Lambda_i,\pi\left(\{t_k\},{\Lambda^{\text{\tiny QW}}}\right)$}
\Comment {Evaluate scores}
\EndFor
\State $best\leftarrow ({\bf Min}(S), \Lambda_{\bf{Min}(S)}$)
\If{$best[0]= 0$}  
\State \Return $best$
\EndIf
\For{$i= 0 \to p_e n_p-1$}
\State HOF$_i\leftarrow$  $(\Lambda_i,S_i)$  sorted by scores
\Comment{Hall of fame}
\EndFor
\State { Insert} HOF { into} $P_{gen+1}$
\For {$j= 0 \to n_p(1-p_e)/2-1$} 
\State $\Lambda_1^j,\Lambda_2^j \leftarrow$ \Call{Tournament}{$P_{gen},S$}
\Comment{Select parents }
\State Add \Call{Crossover}{$\Lambda_1^j,\Lambda_2^j$}  to children
\Comment{Children}
\EndFor
\For {$i= 0\to n_p(1-p_e)-1$}
\State Apply \Call{Mutation}{children$_i$}
\Comment{Mutation}
\EndFor
\State Insert children in $P_{gen+1}$
\State $gen\leftarrow\,gen+1$
\EndWhile 
\end{algorithmic}
\end{algorithm}
The values of the hyperparameters are reported in   Table \ref{tab1}: 

\renewcommand{\arraystretch}{2}
\begin{table}[ht]
\resizebox{0.8\columnwidth}{!}{
\begin{tabular}{llc}
\multicolumn{2}{c|}{Parameter}                             & Value                \\ \hline
$n_g$ & \multicolumn{1}{l|}{Maximum number of generations} & 100                  \\
$n_p$ & \multicolumn{1}{l|}{Population size}               & $2\cdot n_c^2$       \\
$p_e$ & \multicolumn{1}{l|}{Elitist probability}            & $0.02$      \\
$k$   & \multicolumn{1}{l|}{Tournament competitors}        & 6                    \\
$p_c$ & \multicolumn{1}{l|}{Crossover probability}         & 0.85                 \\
$p_m$ & \multicolumn{1}{l|}{Mutation probability}          & 0.05                 \\ \hline
\end{tabular}
}
\label{tabella}
\caption{Genetic algorithm parameters}
\label{tab1}
\end{table}

\subsection{Genetic Operations}
We define the genetic functions which are used in the algorithm: \\
{\it Fitness evaluation}. The algorithm evaluates the fitness of each individual in the population $\Lambda_i$ by evolving the initial state using the couplings in $\Lambda_i$ and measuring the distance between the  generated  and  measured probabilities concatenated at different times $t_k$, i.e. $\pi_x(\{t_k\},\Lambda_i)$ and $\pi_x(\{t_k\},\mathbf{\Lambda^{\!\text{\tiny QW}}}) $ respectively. The distance is measured with the Kullback-Leibler divergence (KLD), defined as:
\begin{equation}
    \text{KLD}(\Lambda_i)=\sum_{x} \pi_x(\{t_k\},\Lambda_i)\log{\left(\frac{\pi_x(\{t_k\},\Lambda_i)}{\pi_x\left(\{t_k\},\mathbf{\Lambda^{\!\text{\tiny QW}}}\right)}\right)}.
\end{equation}
We note we have also tried metrics such as the Kolmogorov distance, obtaining analogous results. \\
{\it Torunament selection.} We select the individuals for reproduction among the whole population running repeated tournaments between $k$ individuals at a time. We need to select $n_p(1-p_e)$ individuals so that, since every couple will produce two children with probability $p_c$, the size of the population remains unchanged at each iteration. During each tournament, $k$ individuals at random are selected among the whole population. The fittest one among the $k$ (i.e. that with the smallest KLD) is chosen as a parent. \\
{\it Crossover.} Children are created two at a time. Both are initialized with the chromosome of one of their parents each. With a probability $p_c$, their chromosomes are crossed over. If the crossover happens, a random integer number smaller than $n_c$ is selected, and serves as the splitting point for the chromosome of the two parents: one child's chromosome will be comprised of the chromosome of the first parent up to the splitting point, and of the second parent thereafter - and viceversa for the other child. \\
{\it Mutation} For each child, each gene can undergo a mutation with a probability $p_m$.  This is achieved by selecting a random number between 0 and 1. If the number is smaller than $p_m$, then the gene will be flipped.

\begin{figure*}[h]
    \includegraphics[width=1\textwidth]{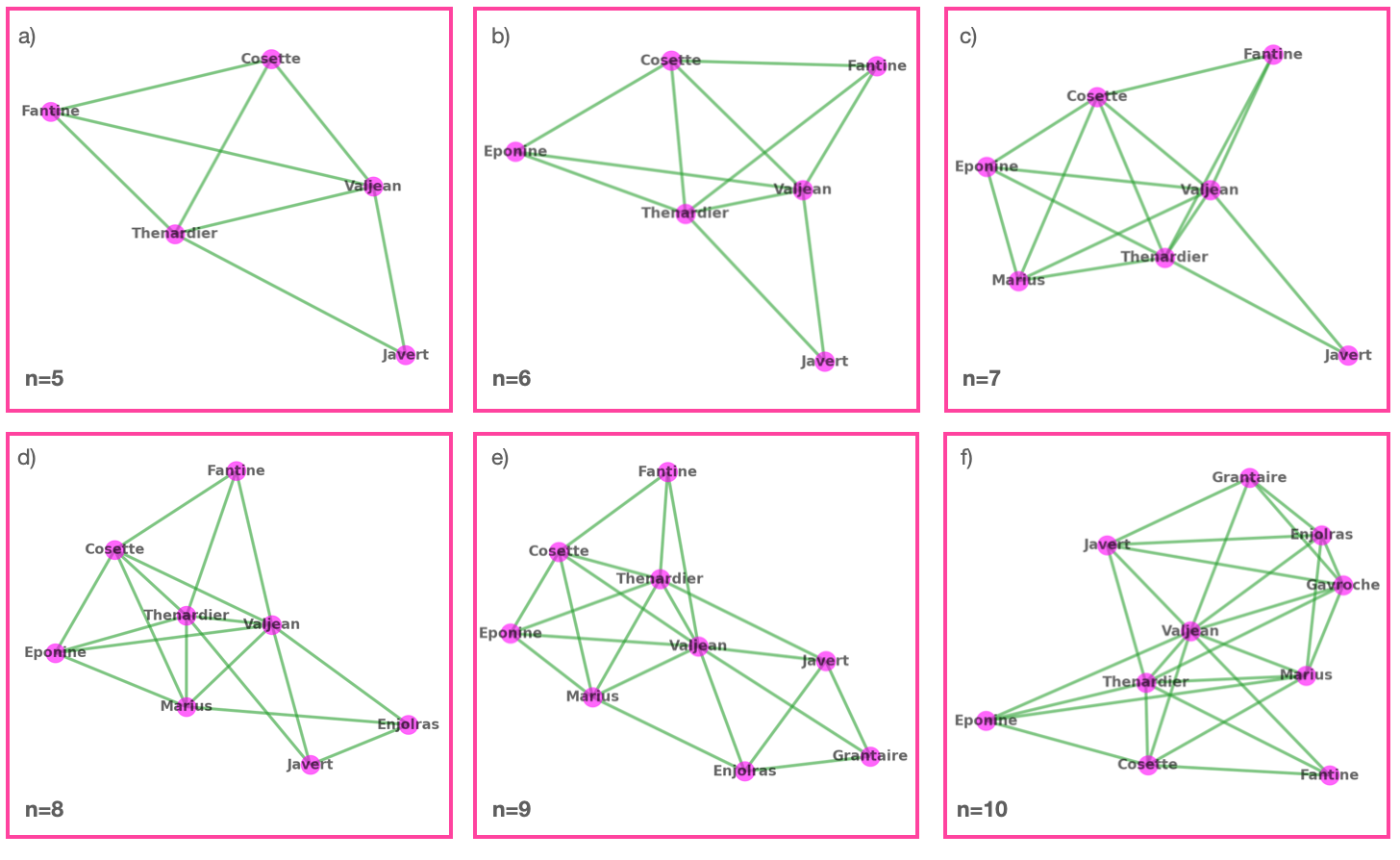}
    \caption{{\bf Random Graph.} Composition of the random graph for n=5-10} 
    \label{lesmis}
\end{figure*}  

The pseudocode for each function is reported in Algorithm \ref{alg2}. 

\begin{algorithm}[H]
\caption{Genetic functions}
\label{alg2}
\begin{algorithmic}[1]
\Function{Fitness}{$\Lambda_i,\pi(t_k,\mathbf{\Lambda^{\!\text{\tiny QW}}}$)}:
  \State Evaluate $\pi(t_k,\Lambda_i$)
  \State Evaluate KLD$(\pi(t_k,\Lambda_i),\pi(t_k,\mathbf{\Lambda^{\!\text{\tiny QW}}}) )$
  \State \Return KLD
\EndFunction
        \Statex
\Function{Tournament}
 {$P_{gen},S$}:
 \State id$\leftarrow$ random integer in $[0,n_p]$
 \For{$j = 0 \to k-2$}
 \State aux$\leftarrow$ random integer in $[0,n_p)$
 \If{$S[\text{aux}]<S[{\text{id}}]$}
 \State id$\leftarrow$ aux
\EndIf
\EndFor
\State \Return $\Lambda[{\text{id}}]$
 \EndFunction
    \Statex
\Function{Crossover}{$\Lambda_1,\Lambda_2$}:
\State Generate a random integer $x$ in $[0,1]$
\If{ $x<p_c$} 
  \State      $y\leftarrow$ random integer in $[0,n_c)$
 \State $\text{child}_1\leftarrow \text{concatenate}(\Lambda_1[0:y],\Lambda_2[y+1:n_c-1])$
 \State $\text{child}_2\leftarrow\text{concatenate}(\Lambda_2[0:y],\Lambda_1[y+1:n_c-1])$
\EndIf
\State \Return { child$_1$,\text{child}$_2$}
\EndFunction
\Statex
\Function{Mutation}{child$_i$}:
 \For{$j= 0\to n_c-1$}
 \State Generate random $x$ in [0,1]
    \If{ $x<p_m$} 
    \State {child$_i[j]\leftarrow 1-$child$_i[j]$}
    \EndIf
    \EndFor
\State\Return{child$_i$}
\EndFunction
\end{algorithmic}
\end{algorithm}

\subsection{Les Misérables graph}
In order to test the algorithm on a graph with a random topology we adopt a simplified version of the graph describing the connections between the characters in the novel Les Misérables by V. Hugo. We use only the main characters, and we fix all the coupling strengths to 1. We start from $n=5$ characters, and then increase $n$ by introducing new characters. The resulting graphs are reported in Fig. \ref{lesmis}

 \begin{figure*}[ht]
    \includegraphics[width=1\textwidth]{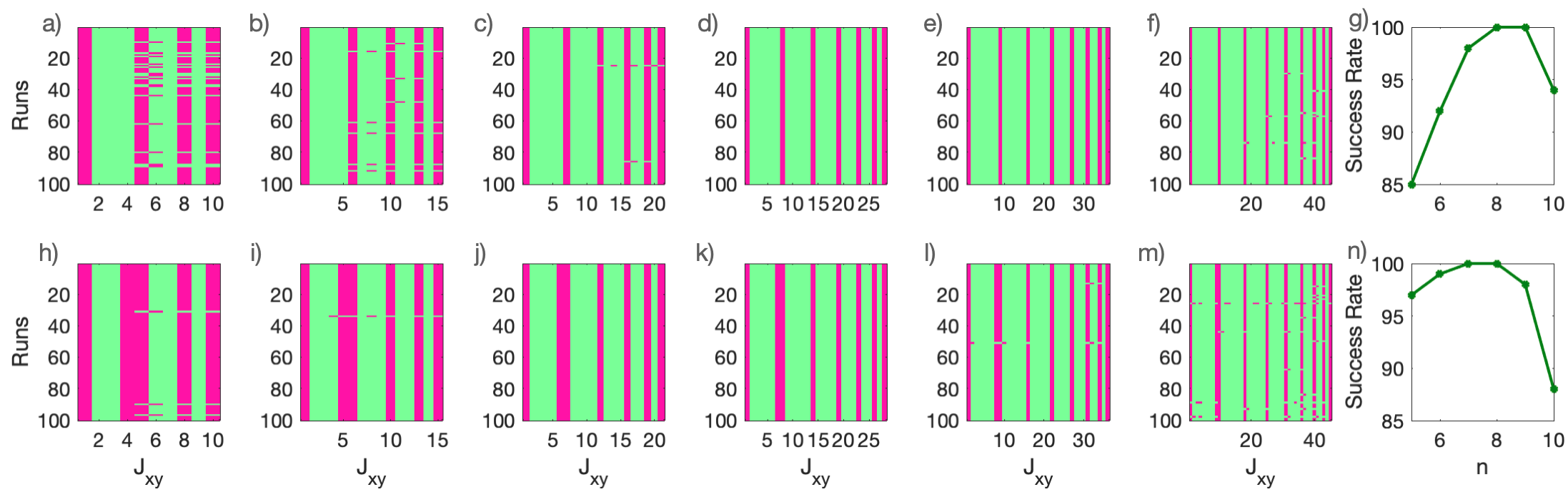}
    \caption{{\bf Results for additional topologies - couplings.} Retrieved couplings over 100 runs for a line (a-f) and circle (h-m) network. g) Success rate for line (g) and circle (n).  } 
    %\label{fig:couplings}
    \label{additional}
\end{figure*}  

\subsection{Results for additional topologies}
\begin{figure}[ht]
    \includegraphics[width=1\columnwidth]{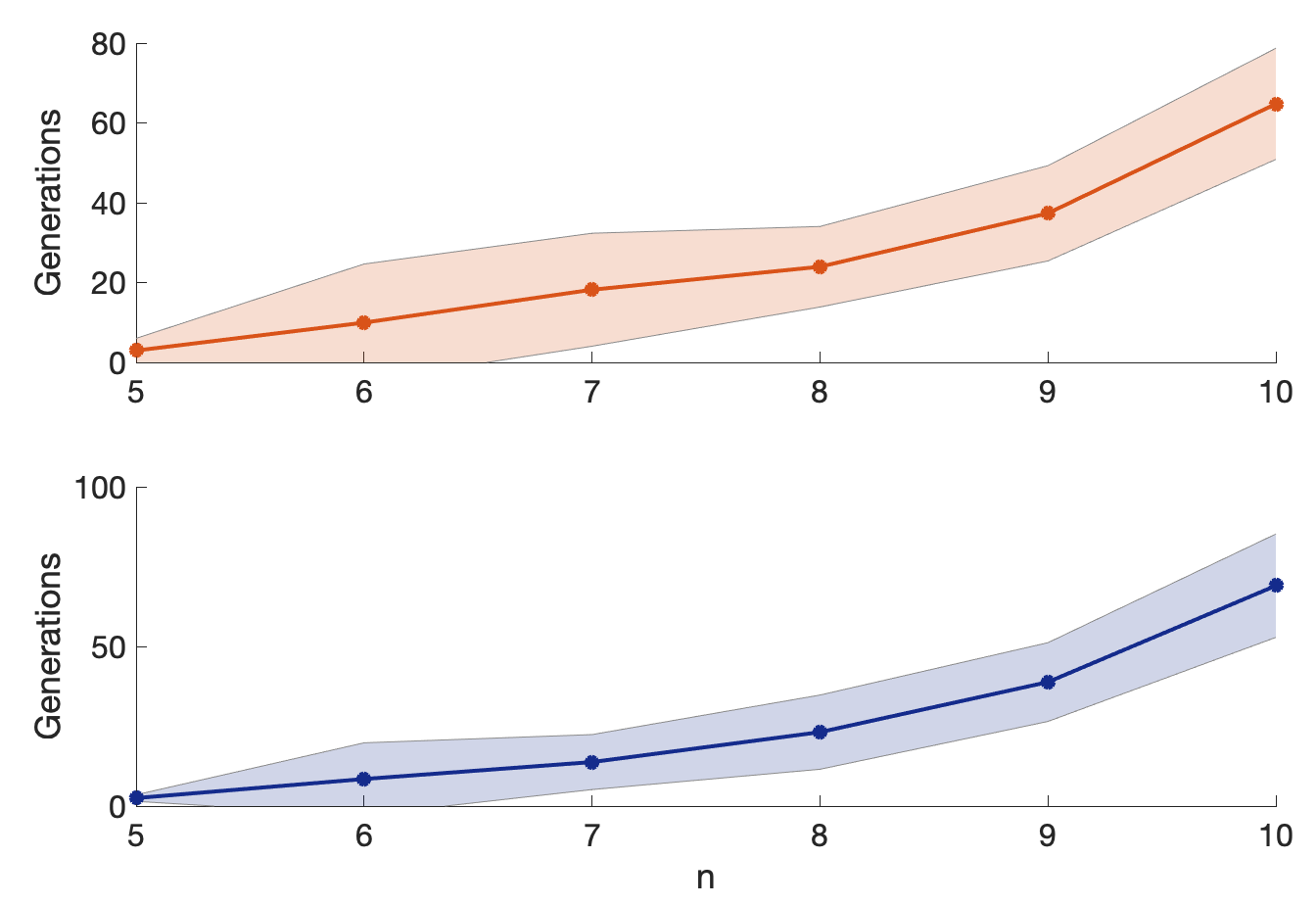}
    \caption{Generations required for convergence for a line network (upper panel) and a circle network (lower panel)}
    \label{additionalgen}
\end{figure}

In Fig. \ref{additional} we report the results obtained without noise for the line and circle topologies.  Panels (a-f) and (h-m) show the couplings for the N=100 runs of the algorithm, while panels (g,n) show the success rate. 
In Fig. \ref{additionalgen} we report the generations needed for convergence.

\subsection{Complete Network with n=10}
\begin{figure}[ht]
    \includegraphics[width=1\columnwidth]{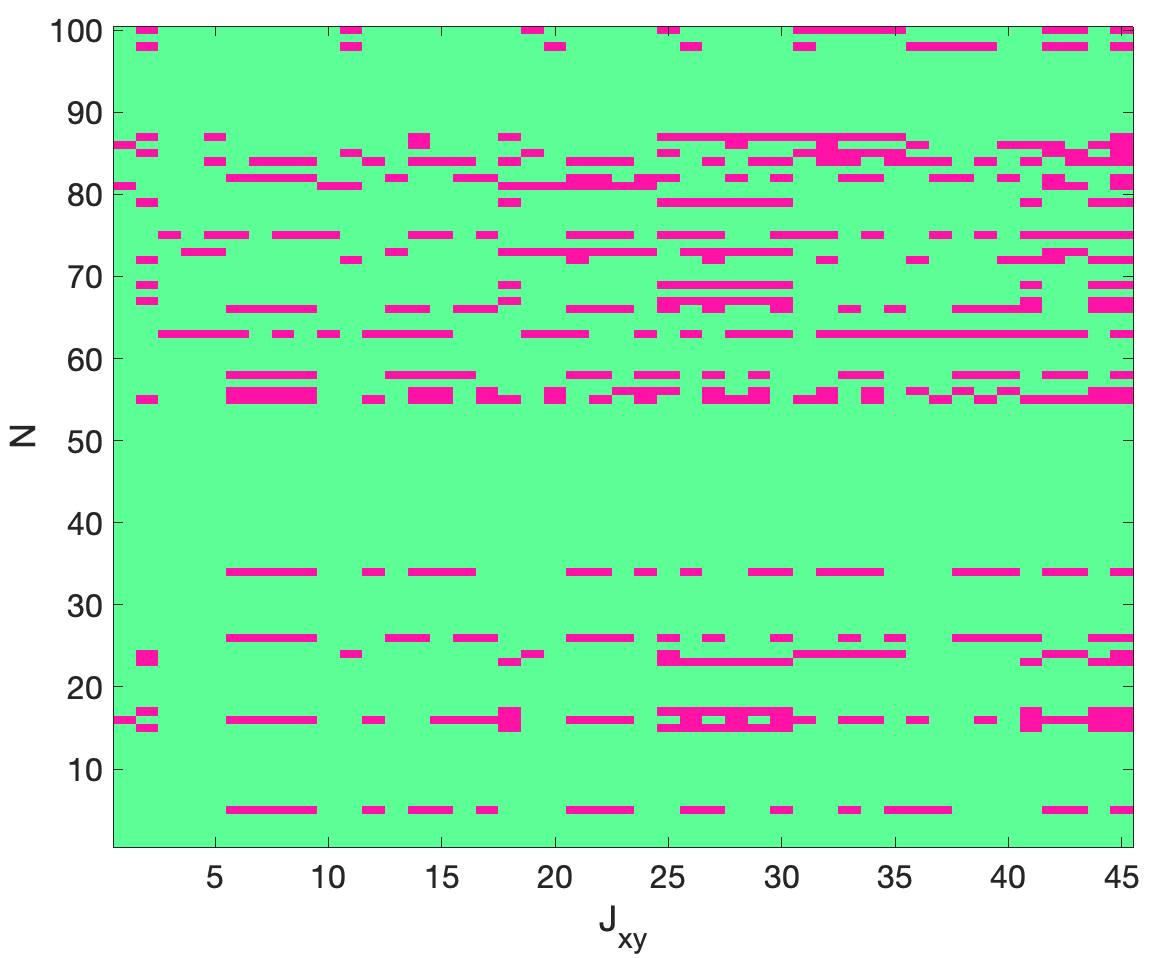}
    \caption{Retrieved couplings for a n=10 complete network using probability distribution measured at three evolution times: $t_1=0.5$, $t_2=0.6$, and $t_3=1$}
    \label{fig:coupling10}
\end{figure}

As shown in the main text, the complete network for n=10 is the most affected by local minima, which prevent the algorithm to converge to the correct solution dropping the success rate to $31\%$. This is because at the chosen times, there are configurations leading to similar probabilities than the complete one. However, it is sufficient to repeat the algorithm adding a third probability measured at $t_3=1$, to drastically increase the success rate up to $73\%$. The retrieved couplings are reported in Fig. \ref{fig:coupling10}.

\subsection{Results with noise for additional resources}

\begin{figure}[ht]
    \includegraphics[width=1\columnwidth]{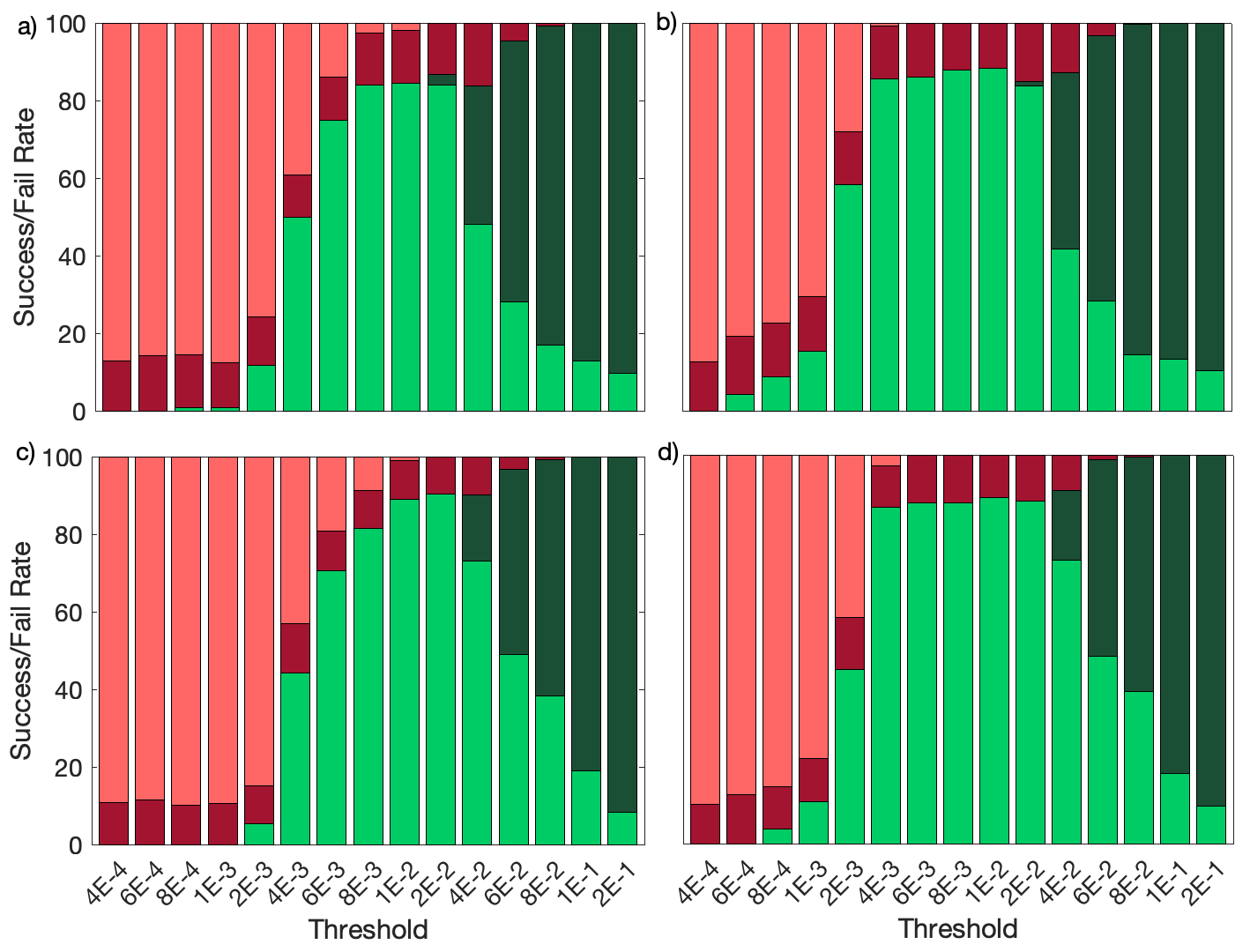}
    \caption{(a-b) Results for star network, (c-d) results for complete graph with $N_r=1000$ (a-c) and $N_r=2500$ (b-d). Light red: false negatives, Dark red: true negatives, Light green: true positives, Dark green: false positives}
    \label{fig:noiseadditional}
\end{figure}

We report additional results with noise for $N_r=1000$, and  $N_r=2500$ for the star and the complete networks with n=5. 
The success/fail rates are shown in Fig. \ref{fig:noiseadditional}

\end{document}